%% file: paper.tex
\documentclass[twocolumn]{bytedance_seed}

\usepackage[toc,page,header]{appendix}

\usepackage{graphicx}
\usepackage{minitoc}
\usepackage{bbding}
\usepackage{stfloats}

\usepackage{microtype}
\usepackage{booktabs} %

\usepackage{hyperref}
\usepackage{tabularx}
\usepackage{pifont}

\newcommand{\charon}{Charon}

\usepackage{subcaption}

\title{Charon: A Unified and Fine-Grained Simulator for Large-Scale LLM Training and Inference}

\author[1,2,*, \dagger]{Mengtian Yang}
\author[1]{Zhekun Zhang}
\author[1]{Mingheng Wu}
\author[1]{Jianwen Yan}
\author[1]{Hanshi Sun}
\author[1, \dagger]{Li-wen Chang}

\affiliation[1]{ByteDance Seed}
\affiliation[2]{University of Texas at Austin}

\contribution[*]{Work done at ByteDance Seed}
\contribution[\dagger]{Corresponding authors}

\abstract{
Deploying large-scale LLM training and inference with optimal performance is exceptionally challenging due to a complex design space of parallelism strategies, system optimizations, and hardware configurations. Accurate and rapid performance simulation is critical for guiding optimization efforts and system studies by validating ``what-if'' hypotheses. To address this, we introduce \charon, a unified, modular, and fine-grained simulator for accurately predicting LLM performance. Experiments show \charon~achieves high accuracy across different models and configurations, with an overall prediction error consistently under 5.35\%, and even under 3.74\% for training with a large-scale GPU cluster. In a practical inference deployment case, \charon~discovered a configuration that improved system throughput over an engineering-tuned baseline, demonstrating its significant real-world value.
}

\date{May 16, 2026}
\correspondence{\email{mtyang@bytedance.com} and \email{liwen.chang@bytedance.com}}

\begin{document}
\maketitle

%不需要目录就注释掉 注意目录不要和第一页放在一块 要有\newpage
%\newpage
%\tableofcontents
%\newpage

\input{sections/introduction}

\input{sections/bg_motivation}

\input{sections/methodology}

\input{sections/experiments}
\input{sections/casestudy}
\input{sections/conslusion}
\input{sections/ackownledgments}

\clearpage

\bibliographystyle{plainnat}
\bibliography{main}

\clearpage

% \beginappendix

% \input{sections/appendix}

\end{document}

%% file: sections/introduction.tex
\section{Introduction}

In recent years, large language model (LLM) workloads have rapidly transitioned from research prototypes to production systems powering search engines, conversational agents, and code assistants. This shift has brought an unprecedented surge in computational demand—for instance, training a model like LLaMA-3 405B can consume over 30 million GPU hours across 16,000 H100 GPUs~\cite{meta-llama/Llama-3-405B}. At this scale, the efficiency of both training and inference depends on the careful coordination of parallelism strategies, network topology, and model architecture. Even small misconfigurations can result in days of wasted compute or intolerable serving latency, leading to substantial financial and operational costs. To address these, performance simulators have been widely explored to capture the behavior of LLM training and serving pipelines, facilitating scalability analysis and hardware–software co-design~\cite{ASTRA-sim, SimAI, Lumos, Echo, LLMServingSim, Vidur}.

\begin{figure}[h]
    \centering
    \vspace{-4pt}
    \includegraphics[width=0.97\linewidth]{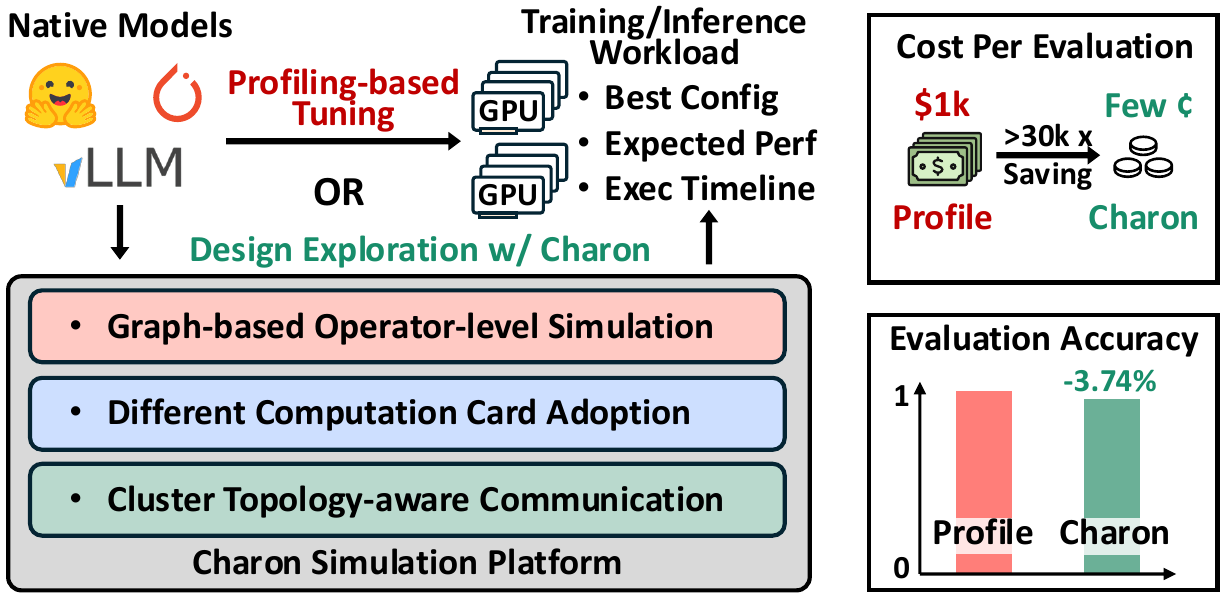}
    \vspace{-3pt}
    \caption{\charon~enables end-to-end, operator-level simulation for LLM training and inference. It delivers more than 30k cost reduction compared with cluster profiling for large-scale experiment, with only 3.74\% total training time error on a large-scale training task. }
    \vspace{-3pt}
    \label{fig:hooker}
\end{figure}

However, existing LLM simulators remain fragmented and insufficient for end-to-end analyses. Most are specialized for either training~\cite{SimAI, Lumos, Echo, dPRO} or inference~\cite{Vidur, LLMServingSim}, forcing engineers to rely on separate, often incompatible tools. They also impose substantial usability overhead, requiring manually building models inside the simulator~\cite{ASTRA-sim, SimAI, Vidur} or pre-processing~\cite{Lumos, Echo} rather than supporting direct use of native models.
Architecturally, existing simulators either lack operator-level granularity in modeling both computation and communication~\cite{ASTRA-sim, SimAI}, or lack the flexibility to modify operator graphs for simulating system or compiler optimizations~\cite{Echo, Vidur}. These limitations hinder fine-grained performance analysis, seamless hardware integration, and optimization exploration.
Table~\ref{tab:intro_exist_sim} summarizes these limitations, including gaps in workload coverage, parallelism support, simulation granularity, and backend flexibility. Consequently, an ideal simulator should support both training and inference, allow direct use of native models, provide operator-level insights, and enable flexible integration of hardware backends and parallelism strategies, allowing accurate design-space exploration and performance analysis at scale.

\begin{table*}[t]
\caption{Capabilities of existing training and inference simulators}
\vspace{6pt}
\resizebox{\linewidth}{!}{ 

\begin{tabular}{|m{3.1cm}<{\raggedright}|m{1.5cm}<{\centering} m{1.7cm}<{\centering} m{1.5cm}<{\centering} m{1.7cm}<{\centering} m{1.5cm}<{\centering} m{2.2cm}<{\centering} m{4cm}<{\centering}|}

\hline
\textbf{}                            & Astra-sim    & SimAI              & Lumos                     & Echo                     & Vidur        & LLMServingSim & \textbf{Charon (This Work)}                  \\ \hline
Training                             & \Checkmark            & \Checkmark                  & \Checkmark                         & \Checkmark                        &  \XSolidBrush            &  \XSolidBrush             & \textbf{\Checkmark}                              \\ \hline
Inference                            & \Checkmark            &  \XSolidBrush                  &  \XSolidBrush                         &  \XSolidBrush                        & \Checkmark            & \Checkmark             & \textbf{\Checkmark}                              \\ \hline
Design Searching                     &  \XSolidBrush            &  \XSolidBrush                  &  \XSolidBrush                         &  \XSolidBrush                        & \Checkmark            &  \XSolidBrush             & \textbf{\Checkmark}                              \\ \hline
Trace Generation                     &  \XSolidBrush            &  \XSolidBrush                  & \Checkmark                         &  \XSolidBrush                        & \Checkmark            &  \XSolidBrush             & \textbf{\Checkmark(w/  3D Timeline)}              \\ \hline
Optimizations (Fusion, Reorder, ...) &  \XSolidBrush            &  \XSolidBrush                  &  \XSolidBrush                         & \Checkmark                        &  \XSolidBrush            &  \XSolidBrush             & \textbf{\Checkmark}                              \\ \hline
Overlapping Slowdown                 & N/A          & Ratio              & N/A                       & Prediction               & N/A          & N/A           & \textbf{Cluster-aware Modeling}         \\ \hline
Computation Modeling                 & Analytical   & Profile / Analytical & Profile / In-house          & Profile                  & Profile      & Profile / Analytical    & \textbf{Analytical / Profile / Prediction}  \\ \hline
Communication Modeling               & Analytical   & Analytical         & In-house                  & Analytical/  Prediction    & Profile      & Analytical    & \textbf{Analytical / Profile / Prediction}  \\ \hline
Parallelism                          & TP / DP / PP     & TP / DP / PP / EP        & TP / PP / DP                  & TP / PP / DP                 & TP / PP        & TP / PP         & \textbf{TP / PP / DP / EP / SP / ZeRO / DualPipe}   \\ \hline
Input                                & Hand Crafted & Mocked Model       & Profiled Traces & Mocked Traces & Mocked Model & Hand Crafted  & \textbf{Native HuggingFace / PyTorch / vLLM Model} \\ \hline

\end{tabular}
}
\label{tab:intro_exist_sim}
\vspace{12pt}
\end{table*}

To overcome the fragmentation and rigidity of existing LLM simulators, which often trade off fidelity for speed or flexibility for usability, we propose \charon, a unified, modular, and fine-grained simulation framework for large-scale LLM training and inference. The key insight is to treat LLM simulation as a compiler-style transformation process, where each stage incrementally refines the model, schedule, and system representation to balance speed, accuracy, and scalability. Architecturally, this perspective allows 
\charon~to have a higher coverage and accuracy than prior simulators by: (i) native model interface directly accepts HuggingFace, vLLM or custom PyTorch models, eliminating hand-crafting or preprocessing; (ii) modular pass-based design which supports plug-and-play analysis and optimization passes, allowing flexible parallelism strategies and new optimization to be modeled without refactoring core infrastructure; (iii) multi-granularity analysis produces system-level aspects as well as fine-grained PyTorch style traces, exposing rich information for performance debugging; (iv) hybrid operator simulation which combined analytical, profiling, and prediction backends, achieving optimal trade off between speed and accuracy.

Empirically, we conducted extensive evaluations of \charon, validating its accuracy against both ground-truth hardware measurements and leading simulators. Across diverse models, including LLaMA3-8B, Qwen3-8B, and Qwen3-30B-A3B, \charon~consistently achieves the highest end-to-end simulation fidelity, with overall prediction errors within 5.35\% of physical hardware and even within 3.74\% with a large-scale GPU cluster. Fine-grained operator-level breakdowns further confirm that \charon~accurately models both computation and communication, including overlapping operators and multi-parallelism strategies. Beyond accuracy, \charon~demonstrates practical impact in design-space exploration and inference optimization: in a case study on LLaMA-3 70B, it automatically discovered an inference deployment configuration that improved system throughput over a manually-tuned baseline.

%% file: sections/bg_motivation.tex
\section{Background and Motivation}

Modern LLMs are based on a decoder-only transformer architecture shown in Figure~\ref{fig:background}(a) \cite{attentionneed}, composed of stacked blocks with self-attention and a feed-forward network, which is often a Mixture-of-Experts (MoE) for parameter efficiency\cite{switchtransformersscalingtrillion}. As shown in Figure~\ref{fig:background}(b), their execution workflows differ: training involves iterative forward, backward, and optimizer passes, while inference operates a forward pass accelerated by the key-value (KV) cache. The scale of these models necessitates both sophisticated parallelism strategies to distribute the workload and extensive performance tuning to optimize execution on large clusters.

\subsection{Parallelism Strategy}
\begin{figure}[h]
    \centering
    \includegraphics[width=\linewidth]{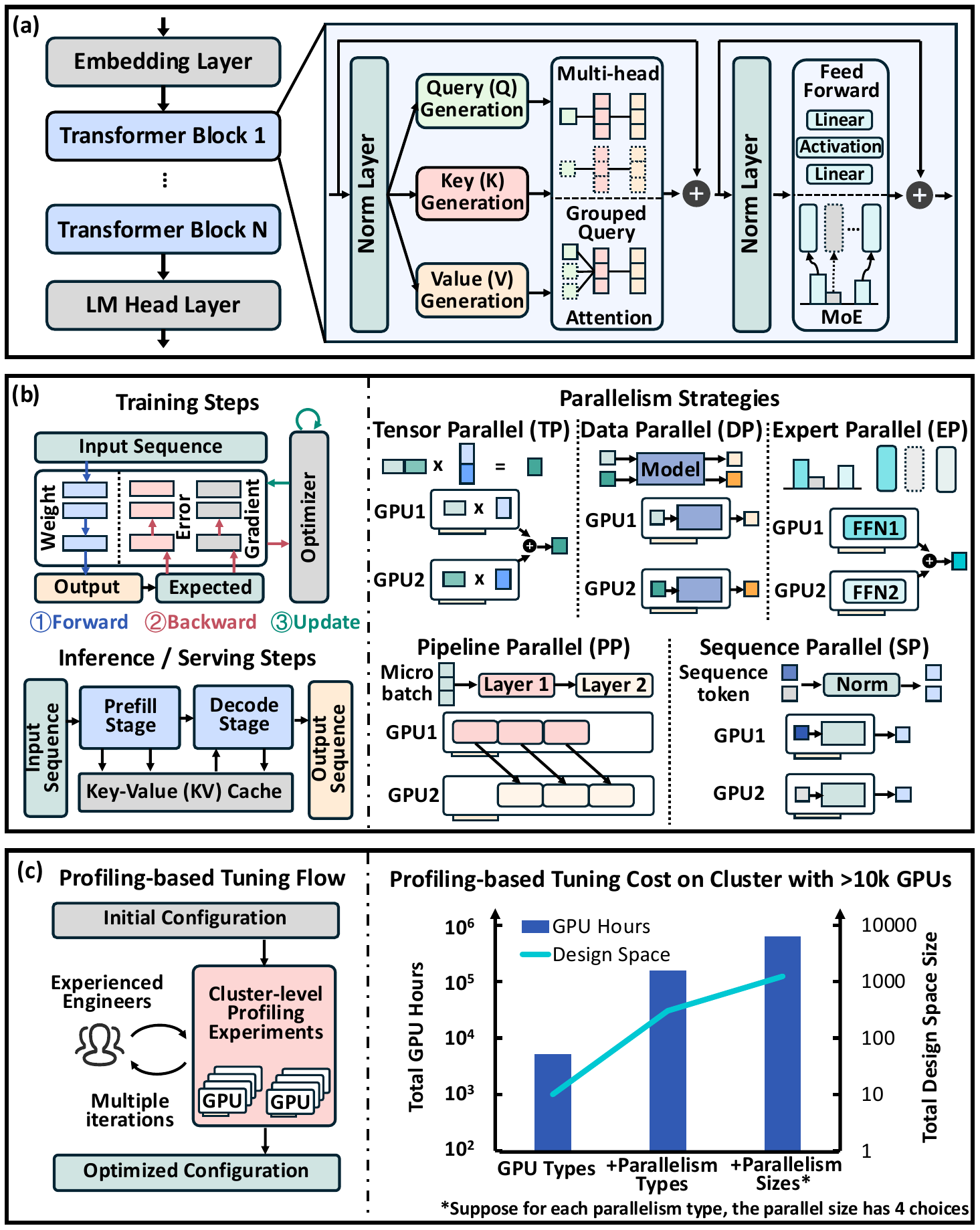}
    \caption{LLM architecture, execution workflows, and tuning challenges: (a) Decoder-only Transformer stack with QKV attention and optional MoE in the feed-forward layers.(b) Training (forward, backward, update) and inference (prefill, decode with KV cache) workflows alongside key parallelism schemes. (c) Iterative profiling-based tuning loop and its GPU-hour/design-space explosion on large GPU clusters.}
    \label{fig:background}
\end{figure}

Due to the enormous LLM model sizes and data involved, training and inference of LLMs almost invariably require large GPU clusters. Consequently, parallelism strategies are critical to achieving acceptable performance. Figure~\ref{fig:background}(b) right part shows the most widely used parallelism strategies: Tensor parallelism (TP) partitions individual weight tensors across multiple GPUs, allowing for wide hidden dimensions at the expense of additional all‐reduce communication during each layer. Data parallelism (DP) traditionally replicates the full model on each GPU and splits the input mini‐batch, synchronizing gradients across replicas after the backward pass. Modern variants such as ZeRO~\cite{ZeRO} and FSDP~\cite{FSDP} divide optimizer states, gradients, and parameters themselves across devices to optimize per‐GPU memory usage and communication volume. Expert parallelism (EP) applies specifically to mixture‐of‐experts models, distributing different expert sub‐networks across GPUs and incurring routing and load‐balancing overhead. Pipeline parallelism (PP) divides the transformer blocks into sequential stages, assigning each stage to a different GPU. Micro‐batches flow through these stages to keep all GPUs utilized. Sequence parallelism (SP) slices the input token sequence across devices for normalization layers, accelerating normalization calculation at the cost of additional communication. Modern deployments typically mix two to four of these schemes simultaneously\cite{megatron,eff_large_scale_training, usingdeepspeedmegatrontrain, sequenceparallelismlongsequence}. For example, TP is used for GPUs in a node, PP is used for inter-node, and EP is used for MoE routing. Each combination imposes distinct demands on computation resources as well as intra-node and inter-node connections, making the design choice very hard for software engineers.

\subsection{Tuning at Cluster Scale}
Considering the scale of modern LLMs, tuning configurations to achieve optimal training and inference performance is essential before deploying models to production clusters. Profiling tools, such as PyTorch Profiler and NVIDIA Nsight Systems, are widely adopted for diagnosing performance bottlenecks in large-scale LLM workloads. By instrumenting both forward and backward passes, these profilers capture fine-grained metrics including operator execution time, GPU SM and memory utilization, kernel-launch overhead, and inter-device communication latency. With this information, engineers can identify bottlenecks such as inefficient attention kernels, suboptimal memory accesses, or network contention and determine which configuration parameters (e.g., batch size, fusion settings, or communication overlap) require adjustment.

Despite the detailed visibility offered by modern profiling tools, tuning training and inference configurations at cluster scale remains highly challenging. As shown in Figure \ref{fig:background}(c), conventional tuning workflow typically follows an iterative ``profile–analyze–tune'' way: engineers run workloads, collect traces, interpret results, and manually adjust settings. However, with ever-growing model sizes and emerging optimization techniques, the design space for large LLMs has expanded dramatically. Even restricting the exploration to GPU model selection and parallelism strategies can involve thousands of possible configurations\cite{Accelerating_DSE}. Evaluating a single design point on a large-scale cluster needs repeated runs (e.g., cold launches and multiple warm-ups) and can consume hundreds of GPU hours. The total exploration cost can approach $10^6$ GPU hours, even under a constrained search that evaluates only four parallelism sizes per parallelism type. Consequently, exhaustive profiling-based tuning becomes prohibitively time-consuming and economically infeasible.

\subsection{Related Works}

\begin{figure*}[ht]
    \centering
    \includegraphics[width=\linewidth]{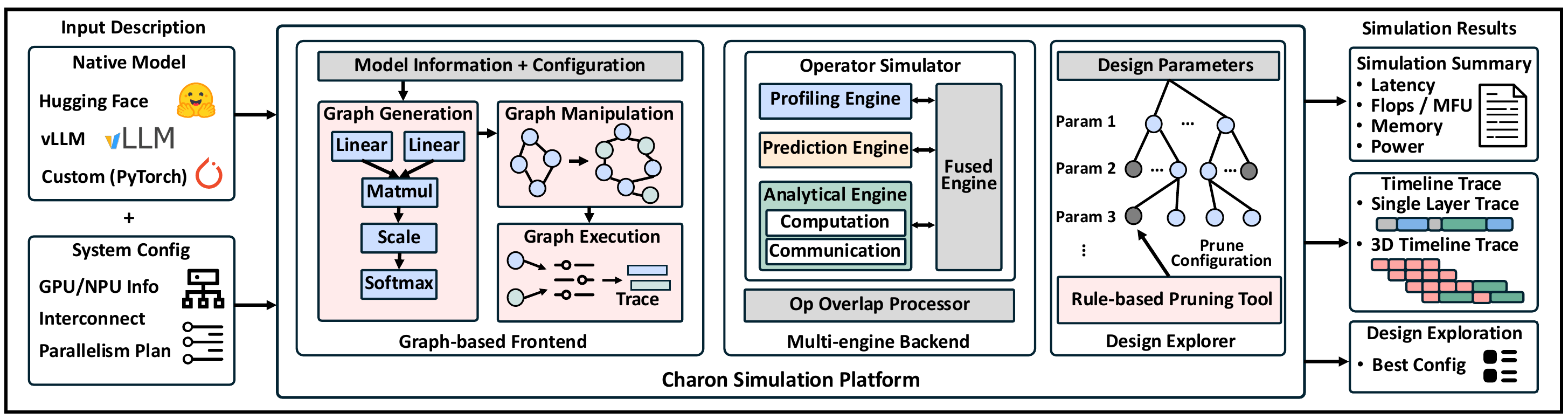}
    \caption{Architecture overview of proposed \charon~simulator. The system consists of a graph-based frontend that constructs forward and backward computation graphs while applying optimizations and analyses; a multi-engine backend that accurately simulates both computation and communication operators; and a design explorer that searches for the optimal configuration through guided pruning. }
    \label{fig:arch_overview}
\end{figure*}

Several simulators have been developed to estimate the performance of LLM training and inference across large-scale systems. ASTRA-Sim~\cite{ASTRA-sim} provides a purely analytical framework for distributed simulation, but requires users to manually construct workload models using predefined templates, limiting ease of use and model fidelity. For training simulation, SimAI~\cite{SimAI} adopts a profiling-based, operator-level performance model for computation while estimating communication at the layer level through either analytical bandwidth models or a slower network-package simulator. Recent work~\cite{SimAI_ext} extends SimAI by adding support for custom device groups and interconnect topologies, yet still depends on mocked models as input. Lumos~\cite{Lumos} and Echo~\cite{Echo} provide operator-level simulation for training but rely on profiling or synthetic traces as workload inputs, introducing additional preprocessing overhead. For inference simulation, Vidur~\cite{Vidur} achieves high accuracy by profiling both computation and communication, but requires models to be rebuilt within its own simulation framework. LLMServingSim~\cite{LLMServingSim} extends ASTRA-Sim to integrate hardware-specific simulators such as NPUs and PIM devices, but still relies on manually crafted workload descriptions.

%% file: sections/methodology.tex
\section{Methodology}
\subsection{Architecture Overview}

Figure~\ref{fig:arch_overview} presents the high-level architecture of \charon, our unified ``all-in-one'' simulator for LLM training, inference and design-space exploration. \charon~ingests native models such as HuggingFace, vLLM, or customized PyTorch models, alongside a system configuration describing GPU or even NPU specifications, interconnect topology, and desired parallelism schemes. From these inputs, \charon~produces both high-level summaries (e.g., FLOPs, model-FLOPs utilization, memory footprint, and power or TDP estimates) and, if enabled, fine-grained execution traces in the style of the PyTorch Profiler (including single-layer timelines and full 3D multi-GPU traces).

\charon's architecture comprises three key components. The frontend parses the model graph and applies a sequence of compiler-style passes—tracing operators, injecting parallelism, scheduling execution, and analyzing results. The backend executes operator-level simulations via interchangeable workers for profiling-based, analytical-based, and prediction-based simulations. It also features an overlap processor to capture communication-computation and communication-communication overlapping and estimate the slowdown for overlapped operators. Finally, the parameter searcher explores the configuration space with suboptimal settings pruning to identify optimal designs regarding cost and performance.

\subsection{Graph-based Frontend}
\begin{figure}[t]
    \centering
    \includegraphics[width=\linewidth]{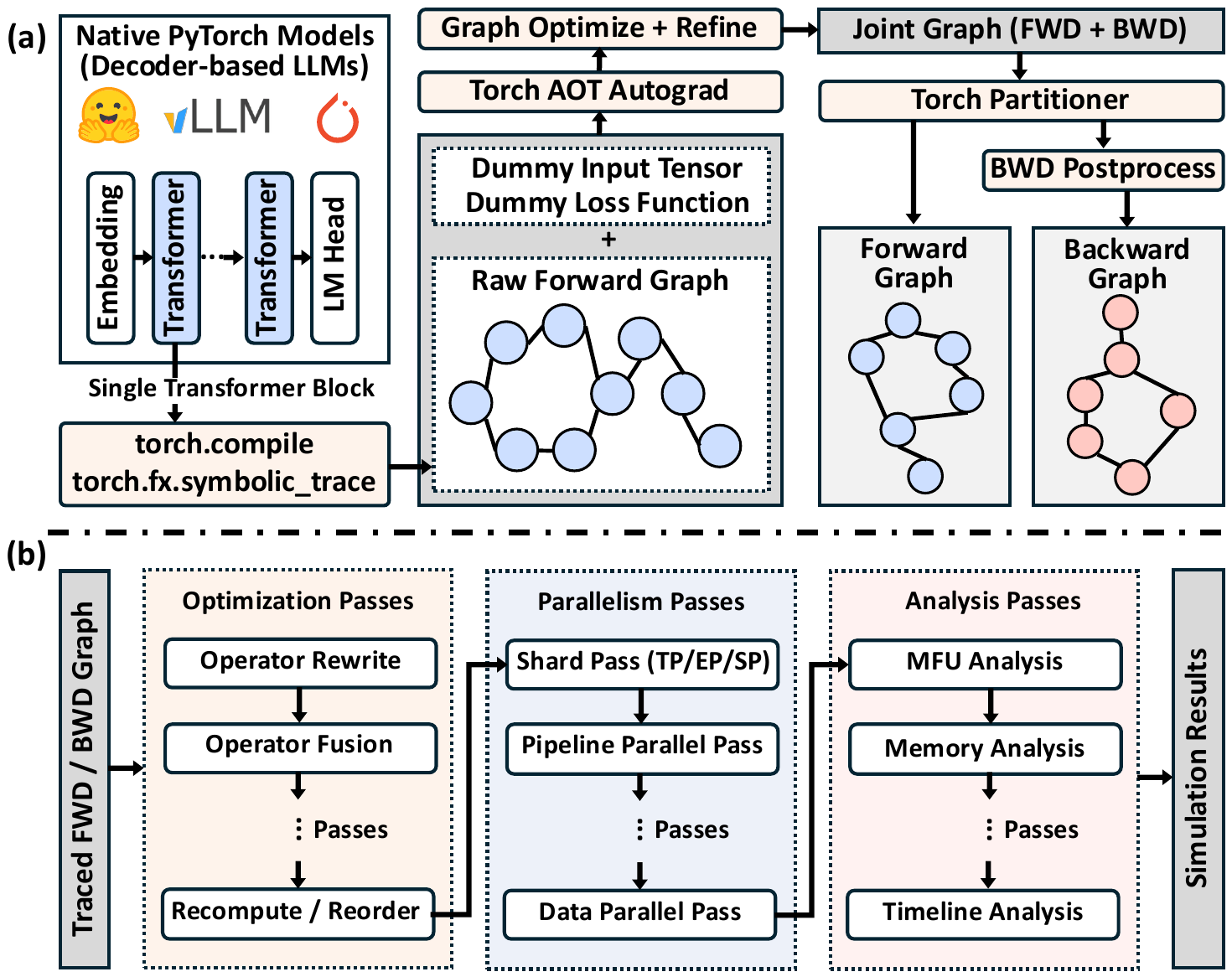}
    \caption{Frontend architecture of proposed \charon~simulator. (a) Trace and generate forward and backward groups from native PyTorch models. (b) Apply compiler-style passes for optimizations and analyses. }
    \label{fig:arch_frontend}
\end{figure}

\charon's frontend is the interface between user's input and the operator-level backend simulator with a computational graph-oriented design. As Figure~\ref{fig:arch_frontend} illustrates, the frontend turns the simulation input into a computational graph and applies multiple graph manipulations to simulate the optimization or analyze the results.

\textbf{(a) Graph Generation with Native PyTorch Model:} 
To maximize usability and support a wide variety of model architectures, \charon~is designed to directly accept native PyTorch models as its simulation input. Since modern LLM architectures typically comprise multiple duplicated Transformer blocks, \charon~extracts and simulates only a single transformer decoder block. This approach substantially accelerates simulation while preserving architectural fidelity and numerical accuracy. 
However, this single-block simulation is a performance optimization primarily for symmetric architectures. For asymmetric models or workloads requiring PP, \charon~traces distinct layers into separate FX graphs according to the model definition and schedules them explicitly per rank, ensuring that all PP ranks and inter-stage dependencies are accurately modeled. Furthermore, to enable disaggregated serving support, \charon~can trace prefill and decode operations into independent computation graphs. This allows simulation of heterogeneous execution where different stages are mapped to different hardware clusters.

The input model can originate from widely used frameworks such as Hugging Face or vLLM, or from any customized PyTorch implementation. This flexibility is enabled by the Graph Tracer shown in Figure~\ref{fig:arch_frontend}(a), which automatically converts the native model into an intermediate computation graph using \verb|torch.fx.symbolic_trace| or \verb|torch.compile|. Specifically, the Graph Tracer first symbolically traces the model to capture its operator-level structure, then leverages the compilation pipeline in \verb|torch.compile| to optimize and lower the traced module into a unified forward compute graph suitable for simulation.

For training tasks that also need the backward computational graph, the graph tracer is designed to automatically generate the backward graph utilizing the \verb|torch._function.aot_autograd| feature. To achieve this, a dummy loss function as well as a fake input tensor will be added to the forward graph to allow dynamic output shapes. After tracing, the joint graph is processed through several passes to enhance its structure and semantics. These include adding tensor metadata, renaming input and weight nodes, and removing useless operations such as view, detach, and other operations that do not alter the tensor's shape or data type. The joint graph is then partitioned into separate forward and backward graph modules using the \verb|default_partition| function. The backward graph undergoes further refinement, including decomposing auto-functionalized operations, removing self-clone operations, and eliminating dead code. Through this series of well-orchestrated steps, graph tracer efficiently and automatically generates a clean and optimized backward graph for the given forward computation.

\textbf{(b) Graph-based Optimization and Parallelism:}
Modern LLMs usually contain multiple optimizations with different parallelism strategies during their training and inference flow. In order to support various optimizations and parallelisms as well as prepare for adopting the upcoming techniques in the future, we proposed a compiler-style design where optimizations and parallelisms are abstracted as graph manipulation passes that apply directly to computational graphs illustrated in Figure~\ref{fig:arch_frontend}(b). Adding or removing the pass will enable or disable a specific optimization for simulation tasks, and different passes can be freely combined to enable joint optimizations.

Operator-level optimizations in \charon, such as operator rewrite and fusion, are implemented through a flexible match-and-replace graph manipulation framework. During this optimization pass, \charon~traverses the computation graph to identify target node patterns that meet predefined matching criteria. Once matched, the framework modifies the node types and attributes to apply the intended optimization, such as merging multiple operators into one or replacing inefficient operator forms. This design enables high extensibility since new optimization rules can be easily added by defining custom match patterns and their corresponding transformation actions. Moreover, for quantization, \charon~can either add a quantization pass to change the precision of the node or directly trace quantized models, enhancing usability and coverage given the widespread adoption of quantization in modern LLM implementations.

Leveraging its pass-based modular design, \charon~supports multi-parallelism by sequentially applying the corresponding passes to the computation graph, providing flexible composition of hybrid parallelism strategies. As illustrated in Section 2, modern LLMs typically employ multiple forms of parallelism when deployed across multi-GPU systems. These parallelisms introduce collective communication operations to coordinate computation and data movement among devices. To model the resulting overhead, \charon~inserts communication operators into the computation graph through a series of dedicated parallelism passes: 

(i) Shard-based parallelisms such as TP, SP, and EP partition tensors and their associated computation operators across multiple GPUs. \charon~traces the computation graph, adjusts the tensor shapes of the sharded operators, and inserts the corresponding communication operators (e.g., \textit{all\_reduce}, \textit{all\_gather}, \textit{reduce\_scatter}) before and after each sharded operation according to the selected strategy. 

(ii) PP partitions the training or inference workflow into multiple stages, with explicit dependencies across devices. In \charon, this is implemented through a schedule pattern generator that constructs inter-stage dependencies and inserts \textit{send}/\textit{recv} communication operators to model data transfer between stages. \charon~supports both the classic 1F1B schedule and the DualPipe schedule with communication and computation overlapping. The resulting dependency graph and communication events are logged for subsequent timeline analysis and visualization. 

(iii) DP partitions the training dataset across GPUs, allowing each device to compute gradients independently. \charon~supports simulation of distributed data-parallel frameworks such as PyTorch DDP, FSDP, and ZeRO. For DDP, the gradients are synchronized via collective communication operations, which is explicitly modeled by inserting communication operators in \charon. FSDP and ZeRO are further supported in \charon~through additional parameters and/or optimizer state sharding, optimizer state synchronization, and prefetching analysis. 

\textbf{(c) Pass-based Multi-granularity Analysis:} \charon~is designed to provide simulation results at multiple granularities for multiple aspects, including coarse-grained system-level results such as model FLOPs utilization, exposed parallelism communication overhead, memory usage, as well as fine-grained operator-level results such as operator latency, efficiency and profiler-style traces. \charon~supports customizing the analysis flow with pass-based analyzers, making it flexible and easy to add new metrics.

The \charon~analyzer supports both dependency-independent and dependency-aware metrics through unified graph-based processing. For metrics that do not require operator dependencies such as model FLOPs utilization, the analyzer operates directly on the computation graph, computing per-node metrics and aggregating them into system-level results through graph traversal. For dependency-aware metrics, such as detailed timeline traces, the analyzer interacts with the \charon~backend to obtain the start and end times of each operator. These are then refined by the operator scheduler, which accounts for inter-operator dependencies (e.g., pipeline stages) to generate an accurate execution timeline. From this dependency-adjusted timeline, profiler-style traces and latency-related metrics (e.g., block latency, end-to-end latency, and FLOPs utilization) are derived. Moreover, optimization techniques such as operator recomputation and communication–computation overlap should be analyzed both before and after optimization to quantify their effects. For example, FLOPs analysis is performed before operator recomputation to capture accurate model-level compute costs. Since both optimization and analysis are applied directly on computation graphs, \charon~natively supports interleaving them within the same simulation flow, thereby improving efficiency, consistency and flexibility.

Accurate peak memory estimation is critical for reliable LLM simulation, as underestimation can lead to out-of-memory (OOM) errors, and overestimation may result in suboptimal configurations. Peak memory consumption in large-scale training typically arises from optimizer states, model weights and gradients, activation tensors, and temporary buffer allocations. Unlike layer-level simulators which can only estimate memory usage based on static tensor sizes (e.g., optimizer states and weights/gradients), \charon’s graph-based design also enables precise modeling of activation and temporary memory by analyzing the liveness of each tensor during gradient computation. The backward computational graph allows \charon~to perform liveness-based traversal to determine exactly when each intermediate tensor is allocated, used, and freed across the backward pass. Consequently, it can reproduce realistic GPU memory behavior, including temporary tensor reuse and deallocation timing that directly influence the peak memory footprint. This fine-grained graph analysis is essential for evaluating memory-critical training configurations (e.g., ZeRO, FSDP, or selective activation checkpoint strategies), where peak memory is reached during backward computation. By leveraging this operator-level liveness modeling, \charon~achieves high-fidelity GPU memory simulation that cannot be captured by non-graph-based simulators.

\subsection{Multi-engine Driven Backend}
\begin{figure}[t]
    \centering
    \includegraphics[width=\linewidth]{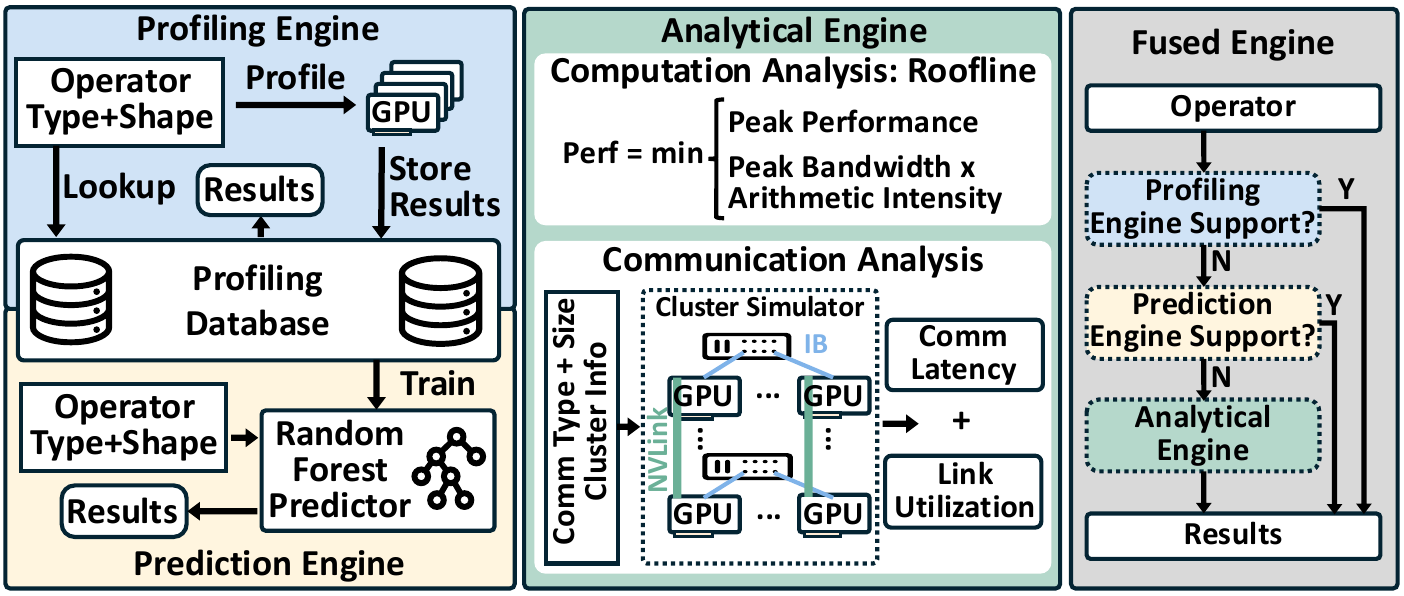}
    \caption{Backend architecture of \charon~simulator. Each operator can be simulated using a profiling, prediction, or analytical engine. A fused backend enables mixed-engine execution, allowing different operators to be simulated by different engines.}
    \label{fig:arch_backend}
\end{figure}

The backend of \charon~is responsible for simulating a single computation or communication operator. As Figure~\ref{fig:arch_backend} shows, \charon~integrates profiling-based, prediction-based, and analytical-based backend engines, as well as a fused backend engine that can use multiple different backend engines to maximize the balance between simulation speed and simulation accuracy.
Additionally, \charon~supports common numerical precision formats like FP32, BF16, FP16, FP8, and INT8 by maintaining precision-specific operator profiles or analytical estimates. Different precision choices impact compute efficiency along with memory usage and communication volume. The simulator explicitly models these factors through precision-aware kernel latency and bandwidth scaling. It also estimates the precision-aware memory footprint for activation, parameter, and temporary buffer sizes.

\textbf{(a) Profiling Engine:} The profiling engine simulates each operator by executing it on the target hardware and profiling runtime, providing the most accurate latency at the cost of higher simulation overhead. For GPU workloads, \charon~automatically generates profiling tasks for each operator, dispatches them to our in-house GPU cluster, and records latency once execution completes. Since profiling every operator instance from scratch is prohibitively expensive, \charon~integrates a profiling database that caches measured results for common operators and input shapes. Upon receiving a simulation request, the profiling engine first queries this database; if a matching operator–shape pair exists, the cached latency is reused directly, thereby significantly improving simulation efficiency.

\textbf{(b) Prediction Engine:} The prediction engine estimates operator latency directly from operator type and input tensor shape using lightweight machine learning models. Its primary goal is to provide fast latency estimation, especially for unseen input shapes not covered in the profiling database. In \charon, each type of operator is associated with a compact random-forest–based predictor trained on data from the profiling database. This design eliminates the need for real-time hardware execution while preserving high accuracy across diverse operator shapes, achieving substantial acceleration in large-scale simulation workloads. 

\textbf{(c) Analytical Engine:} The analytical engine analyzes the execution time of the operator with mathematical modeling according to the operator computation and memory requirement, as well as the hardware capability. For computation operators, \charon's analytical engine utilizes the roofline~\cite{roofline} model, which calculates the computation time and memory access time on targeted hardware and takes whichever is longer as the final computation time for this operator. Hardware FLOPs and memory bandwidth are pre-configured according to the simulation hardware, and operator computation FLOPs/memory accesses are computed on-the-fly according to its input shape.

For communication operators, \charon~employs a hierarchical link-centric model to ensure cross-platform portability and accuracy. Instead of relying on theoretical specifications, \charon~models the cluster topology using calibrated per-hop latency and effective bandwidth derived from profiling. \charon~supports both Ring and Tree collective communication algorithms across diverse topologies (e.g., Ring, Switch, and Mesh). The analytical engine decomposes high-level collective operations into physical link-level data transfers. For each link, the latency is calculated by aggregating the calibrated handshake latency and the transmission latency based on data size and effective bandwidth. This granular approach allows \charon~to precisely evaluate congestion based on bandwidth sharing and topology constraints.

\textbf{(d) Fused Engine:} Fused engine in \charon~is designed to integrate multiple simulation backends within a single execution flow. It enables an adaptive trade-off between simulation speed and accuracy, while maintaining compatibility with emerging models and newly introduced operators. Operators lacking profiling and prediction data can be executed using the analytical engine, while others are simulated using the profiling-based or prediction-based engines. This flexibility is implemented through a prioritized fallback mechanism: each engine maintains a registry of supported operators, and the fused engine dynamically selects the highest-priority backend available for each operator, falling back to lower-priority engines when necessary. This design ensures robust coverage of heterogeneous workloads without sacrificing overall simulation fidelity or scalability.

\subsection{Operator Overlap} 
Modern LLM training and inference frameworks usually overlap communication operators with either compute operators or other communication operators to improve overall performance, which needs to be carefully handled during the simulation to get fine-grained accurate traces. \charon~integrates a coarse-grained ratio-based slowdown model for all overlapped operators, as well as a fine-grained bandwidth-aware model for communication-communication overlap.

The ratio-based slowdown model applies slowdown factors to the overlapped part of two operators. The slowdown factor is engineered from profiling data with targeted hardware clusters. For compute-communication overlap, two separate slowdown factors are used for computation and communication operators. And for communication-communication overlap, the same slowdown factor is shared for two communication operators. The slowdown factor only applies to the portion that the operator has overlapped with other operators.

\begin{figure}[t]
    \centering
    \includegraphics[width=\linewidth]{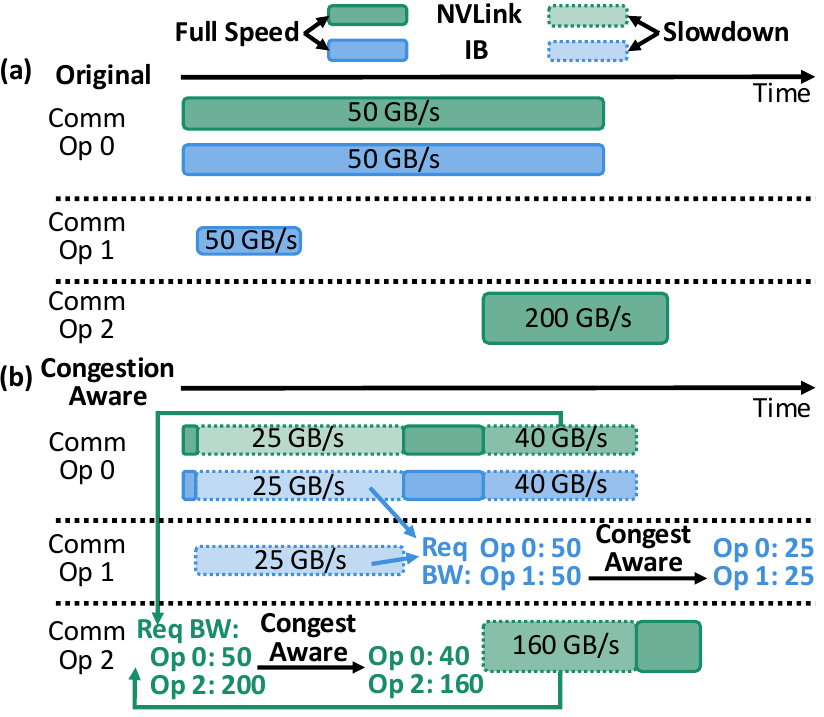}
    \caption{Bandwidth-aware communication operators overlap in the proposed \charon~simulator, (a) shows the timeline of original communication operations, and (b) shows the timeline with congestion-aware simulation.}
    \label{fig:comm_overlap}
\end{figure}

\begin{figure*}[h]
    \centering
    \includegraphics[width=\linewidth]{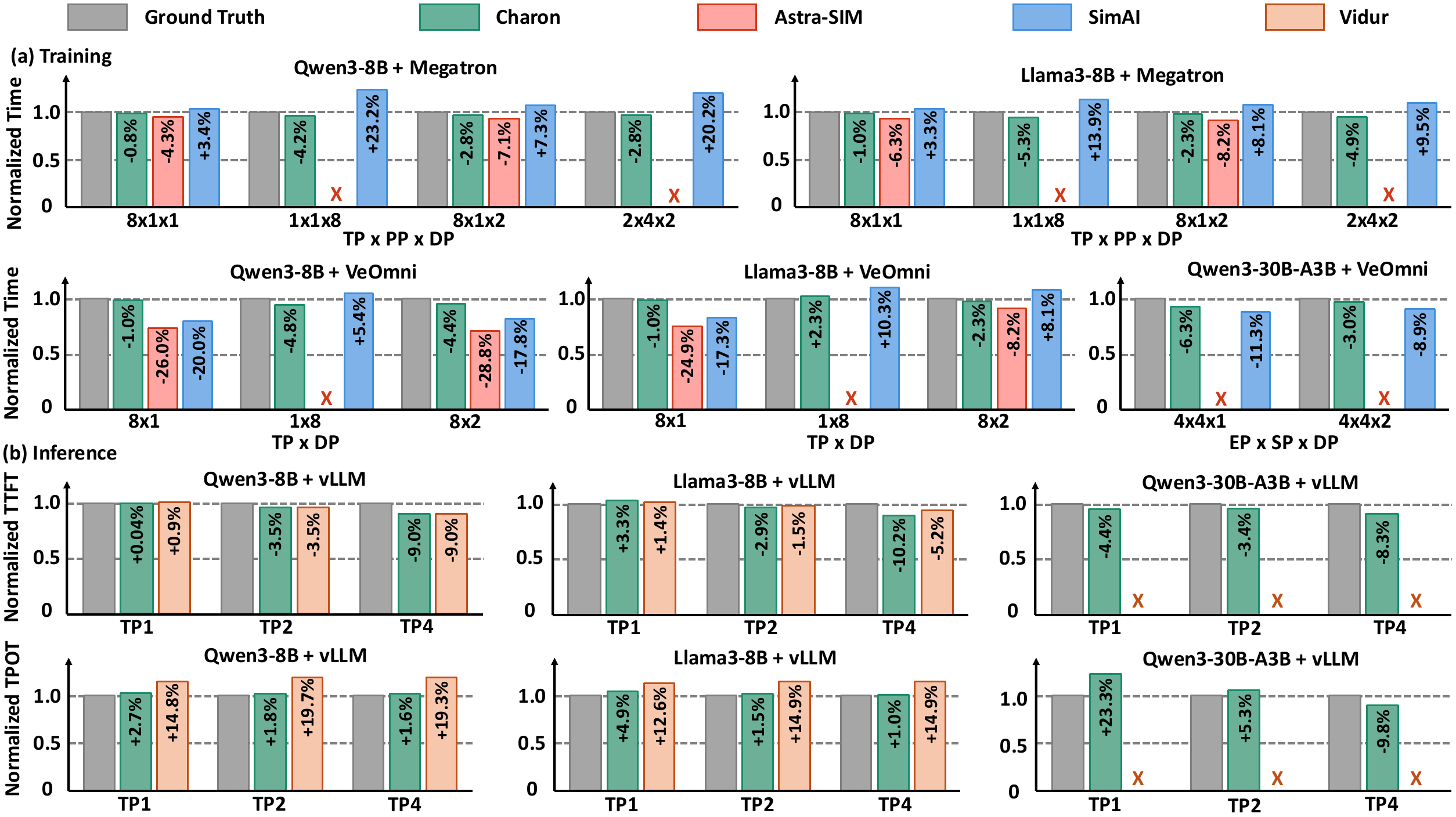}
    \caption{End-to-end time comparison results for \charon~and other simulators against measurement ground truth. ``X'' means the simulator cannot support configurations or cannot give valid results.}
    \label{fig:e2e_training}
\end{figure*}

A fine-grained bandwidth-aware slowdown model is available in \charon~when using analytical engine for the communication-communication overlap. As shown in Figure \ref{fig:comm_overlap}, the slowdown for each operator is decided by the effective bandwidth as well as the link congestion in the cluster. For each portion of the overlapped operators, \charon~checks the link congestion for each interconnect hierarchy and calculates the slowdown according to the effective bandwidth competition ratio to simulate the underlying network packet-level congestion control.

\subsection{Design Space Exploration}
As an LLM training and inference simulation solution, identifying the optimal infrastructure configuration for targeted tasks, such as the number of GPUs, parallelism strategies, and parallelism sizes, emerges as a crucial application in \charon. To eliminate the engineering needed for the analysis of the simulation results in order to use \charon~for design space exploration, \charon~is built in with a native search for the design space.

The design space exploration begins with a targeted model and task, and \charon~will take the entire design space, including different choices for GPU numbers and parallelism sizes from the user's input. To maximize search efficiency, the design exploration tool supports pruning search spaces according to rules. Users can pre-define the known inefficiency cases in \charon~and the design space exploration tools can prune the corresponding sub-spaces by directly skipping the simulation.

%% file: sections/experiments.tex
\section{Experiments}

In this section, we demonstrate the capability and accuracy results of the proposed \charon~simulator.

\subsection{End-to-end Simulation Accuracy}

We evaluated the simulated end-to-end runtime of \charon~against existing LLM simulators as well as the ground truth measurements on both training and inference tasks. To demonstrate \charon’s capabilities, we selected two dense models, Qwen3-8B and LLaMA3-8B, and one MoE model, Qwen3-30B-A3B, as representative case studies. For training simulation, we benchmarked \charon~against Astra-SIM 2.0~\cite{ASTRA-sim} and SimAI~\cite{SimAI}, measuring simulation performance under the Megatron~\cite{megatron} and VeOmni~\cite{veomni} training frameworks. For inference simulation, we compared \charon~with Vidur~\cite{Vidur} by simulating workload within the vLLM\cite{vLLM} framework. All baseline simulators are examined to fix minor bugs and support the new Qwen3 and Llama3 models, as well as re-calibrated according to the profiling results to make a fair comparison.

As shown in Figure \ref{fig:e2e_training}, \charon~achieved the best end-to-end time accuracy among both training and inference tasks with different models and different frameworks. For training tasks, \charon's operation graph-based simulation can accurately simulate the time of each operator level with accurate backends, as well as satisfy the operation of different frameworks. Therefore, the end-to-end time is more accurate than the analytical-based simulator Astra-SIM. For comparison with SimAI, although both use profiling-based backends for computation operators, SimAI simulates the communication based on layer-level information since its communication simulation is based on Astra-SIM 1.0~\cite{astra_sim_1}. This will lead to inaccuracy in handling communication overlaps. For Inference tasks, both \charon~and Vidur can provide accurate time-to-first-token (TTFT), as Vidur uses profiling-based computation and communication simulation. However, for time-per-output-token (TPOT), Vidur cannot produce very accurate results because its predication engine is not accurate enough for small operations.

\begin{table}[t]
    \centering
    \caption{Simulation breakdown for Qwen3-8B. ``Prof'' denotes profiling results and ``Sim'' denotes \charon~simulation results, all results are in microseconds(us).}
    \label{fig:placeholder} % Moved main label right after main caption
    
    % First Subtable
    \begin{subtable}{\linewidth}
        \centering
        \caption{Training breakdown with TP8 on Nvidia Ampera GPU, F indicates forward steps, and B indicates backward steps.}
        \label{tab:training_breakdown}
        \begin{tabular}{l|cc|cc}
        \textbf{Operators} & \multicolumn{1}{l}{\textbf{Prof(F)}} & \multicolumn{1}{l|}{\textbf{Sim(F)}} & \multicolumn{1}{l}{\textbf{Prof(B)}} & \multicolumn{1}{l}{\textbf{Sim(B)}} \\ \hline
        Attention    & 1842                                & 1770                                 & 30275                               & 30329                              \\
        Feed-Forward          & 6589                                & 6490                                 & 40280                                & 38430                               \\
        Others       & 3842                                & 3788                                & 8743                                & 8658                               \\ \hline
        All-Gather           & 13180                                & 12980                                & 13130                                & 12980                               \\
        Reduce-Scatter           & 13876                               & 12980                                & 14500                                 & 12980                              
        \end{tabular}
    \end{subtable}
    
    \vspace{12pt} % Added clean vertical spacing between the two subtables
    
    % Second Subtable
    \begin{subtable}{\linewidth}
        \centering
        \caption{Inference breakdown with TP1 on Nvidia Hopper GPU, P indicates prefill steps, and D indicates decode steps.}
        \label{tab:inference_breakdown}
        \begin{tabular}{l|cc|cc}
        \textbf{Operators} & \multicolumn{1}{l}{\textbf{Prof(P)}} & \multicolumn{1}{l|}{\textbf{Sim(P)}} & \multicolumn{1}{l}{\textbf{Prof(D)}} & \multicolumn{1}{l}{\textbf{Sim(D)}} \\ \hline
        Attention    & 3923                                 & 3906                                 & 58.206                                 & 72.1                              \\
        Feed-Forward           & 9232                                 & 9195                                 & 93.47                               & 109                               \\
        Others       & 141                                  & 142                                  & 6.84                                   & 12                               
        \end{tabular}
    \end{subtable}
\end{table}

\subsection{Time Breakdown}

To demonstrate the fine-grained simulation accuracy of the proposed \charon~simulator, we provide a detailed breakdown of simulation time across individual operations. For training evaluation, we analyze the operation-level simulation of the Qwen3-8B model executed on the VeOmni framework under TP8. For inference evaluation, we report the corresponding breakdown using the Qwen3-8B model in the vLLM framework.

Table~\ref{tab:training_breakdown} presents the breakdown of the training simulation results, while Table~\ref{tab:inference_breakdown} summarizes the corresponding inference simulation results. These results demonstrate that \charon~achieves high simulation accuracy not only at the end-to-end but also at the operator-level granularity.

\begin{figure}[b]
    \centering
    \includegraphics[width=\linewidth]{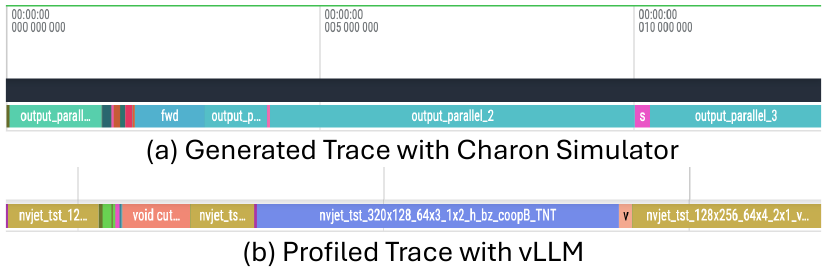}
    \caption{Comparison of simulation traces generated by \charon~and profiled vLLM serving traces. The traces represent a single layer extracted from the full-model simulation and profiling of Qwen3-8B.}
    \label{fig:traces_vs_vllm}
\end{figure}

To further validate the fine-grained accuracy of \charon, we compared the simulation-generated execution traces against hardware-profiled traces. Figure~\ref{fig:traces_vs_vllm} illustrates a single transformer layer trace extracted from a full-model simulation of Qwen3-8B alongside the corresponding profiled trace from vLLM serving. The comparison demonstrates that \charon~accurately simulates operator-level latency and the timeline traces, closely matching the actual hardware execution behavior.

\begin{figure}[t]
    \centering
    \includegraphics[width=\linewidth]{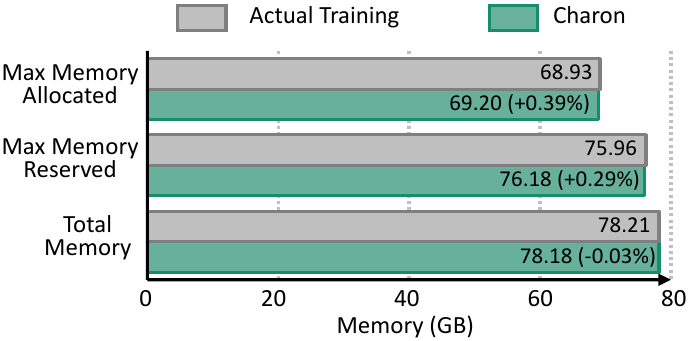}
    \caption{Experimental results for \charon~memory prediction accuracy during Qwen3-30B-A3B MoE model training (FSDP=8, batch\_size=2, seqlen=8192).}
    \label{fig:memory_sim}
\end{figure}

\subsection{Memory Prediction Accuracy}

Beyond execution time, accurate GPU memory estimation is critical for large-scale deployments, particularly for Mixture-of-Experts (MoE) model training, where dynamic routing introduces complex memory access patterns. We validated \charon's memory simulation fidelity during the training of the Qwen3-30B-A3B MoE model on 8 GPUs (configured with FSDP=8, a batch size of 2, and a sequence length of 8192). As shown in Figure~\ref{fig:memory_sim}, \charon~achieves high precision in predicting memory dynamics per GPU. By incorporating calibrated collective communication buffer overheads and dynamic fragmentation effects, the simulation errors for the maximum allocated memory, maximum reserved memory, and total memory footprint are merely +0.39\%, +0.29\%, and -0.03\%, respectively. This confirms \charon's capability to reliably capture realistic memory allocations and provide a faithful representation of the actual memory footprint during complex MoE training.

\subsection{Backend Ablation Evaluation}

\begin{figure}[t]
    \centering
    \includegraphics[width=\linewidth]{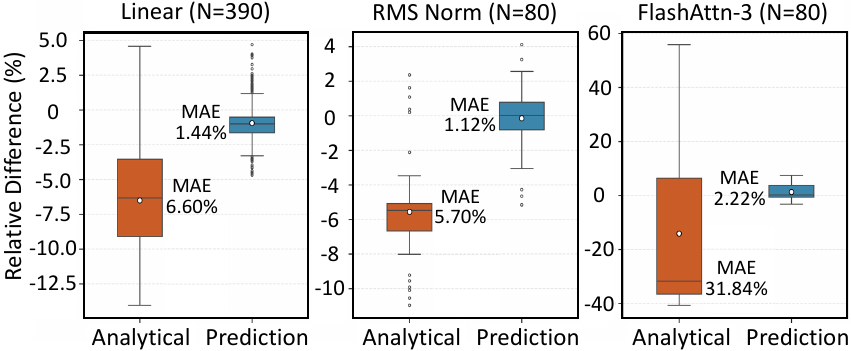}
    \caption{Statistical indicators of analytical engine and prediction engine accuracy deviation for $N$ unseen Linear, RMSNorm, and FlashAttn-3 operators.}
    \label{fig:analytical_vs_prediction}
\end{figure}

To demonstrate the effectiveness of our hybrid multi-engine backend, we conducted an ablation study comparing the simulation accuracy of the analytical engine (Roofline model) versus the prediction engine for unseen tensor shapes. Figure~\ref{fig:analytical_vs_prediction} presents the statistical deviation of prediction accuracy for Linear, RMSNorm, and FlashAttention-3 operators. While the analytical engine provides reasonable estimates for simpler kernels like Linear and RMSNorm, it struggles with complex operators, exhibiting a 31.84\% Mean Absolute Error (MAE) for FlashAttention-3. In contrast, the prediction engine consistently maintains high accuracy across all operators, achieving an MAE of 1.44\%, 1.12\%, and 2.22\%, respectively. This highlights the prediction engine's superior capability in generalizing to complex, unseen workloads without relying on pure analytical approximations.

\subsection{Across Different GPU and Cluster Scale}

We further evaluated \charon~across different GPU architectures as well as both small-scale and large-scale clusters to demonstrate its versatility and scalability. To this end, we selected several in-house profiling experiments targeting both training and inference performance debugging, and configured \charon~to simulate these scenarios. We then compared the end-to-end latency per training or inference step between the profiling measurements and the corresponding simulation results.

\begin{figure}[t]
    \centering
    \includegraphics[width=\linewidth]{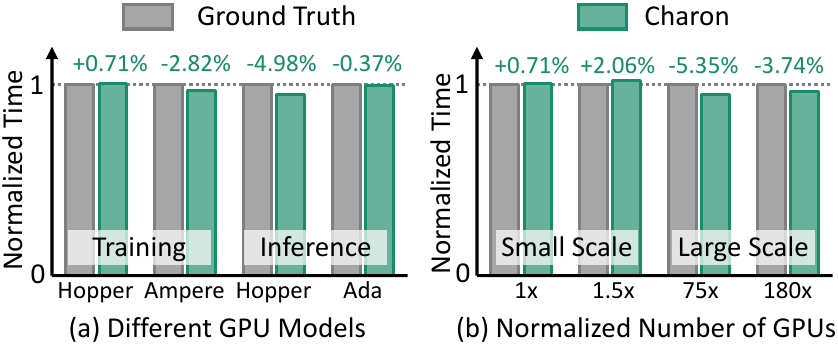}
    \caption{Comparison results between Ground Truth (Profiling) and \charon~for different GPU models under both small-scale and large-scale clusters.}
    \label{fig:diff_gpu_scale}
\end{figure}

Figure~\ref{fig:diff_gpu_scale} reports the normalized end-to-end latency per training and inference step. As shown in Figure~\ref{fig:diff_gpu_scale}(a), \charon~consistently achieves accurate end-to-end simulation across diverse GPU models, covering both Nvidia Hopper / Ampere (H800/A100) training GPUs as well as Nvidia Hopper / Ada Lovelace (H20/L20) inference GPUs and achieving overall simulation errors within 4.98\%. Furthermore, across varying cluster scales in training tasks shown in Figure~\ref{fig:diff_gpu_scale}(b), \charon~demonstrates scalability to extra-large configurations with nearly ten thousand GPUs and the optimal combination of all kinds of parallelism (DP, PP, EP, SP, and TP), while keeping simulation error under 3.74\%. Meanwhile, \charon~consistently maintains the maximum overall error under 5.35\% across both small-scale and large-scale clusters.

The minor prediction error gaps observed across different hardware and scales are primarily due to the inherent runtime variability present in physical execution traces. Factors such as network communication jitter, dynamic congestion, and data-dependent kernel execution randomness introduce stochastic behaviors in real-world clusters. Because \charon~currently does not deterministically model these micro-level stochastic variations, minor deviations between the simulated and profiled results are expected. However, as demonstrated by the consistently low error margins, these gaps do not impact the simulator's ability to provide highly reliable performance and scalability analyses.

%% file: sections/casestudy.tex
\section{Case Study}

In practice, \charon~serves as a versatile platform for conducting comprehensive "what-if" analyses to guide system-level design decisions. Beyond basic configuration tuning, engineers can utilize \charon~to evaluate the impact of complex compilation and execution strategies without requiring a full compiler implementation. For example, operator fusion can be simulated by defining localized pattern replacements with fused operator models, allowing users to assess latency improvements prior to deployment. Similarly, memory-efficient operator reordering and overlapping strategies can be explored directly on the operation-level graph, with the resulting memory footprint and scheduling timelines instantly reflected in the simulation output. Furthermore, \charon's pass-based architecture naturally supports the evaluation of advanced techniques such as activation checkpointing, selective offloading, and heterogeneous device placement.

To illustrate \charon's practical utility in navigating complex design spaces, we present two detailed case studies. First, we demonstrate how \charon~facilitates the design of a novel dynamic Sequence Parallelism strategy by analyzing the fine-grained tradeoffs between computation and communication. Second, we showcase \charon's capability to perform rapid, multi-objective optimization for LLM inference, identifying optimal deployment configurations that balance system throughput against strict user-facing latency constraints.

\subsection{Dynamic Sequence Parallel Strategy}
Sequence parallelism (SP) is commonly used to optimize prefill latency in LLM serving by distributing attention computation across ranks. In this case, zigzag attention divides the sequence dimension into 2xSP chunks, assigning each rank two chunks in a zigzag pattern to balance the workload. This strategy has been widely adopted in large-scale models such as LLaMA-3~\cite{meta-llama/Llama-3-405B}. While zigzag attention improves compute balance compared to naive partitioning, it is not always optimal. For short sequences, zigzag attention is less efficient: it over-partitions the input and adds disproportionate all-gather communication overhead, which outweighs the potential compute savings. To address this, we extend zigzag with dynamic SP, a fine-grained scheme that assigns different SP configurations per request within a batch, ensuring the best end-to-end latency.

\begin{figure}[t]
    \centering
    \includegraphics[width=\linewidth]{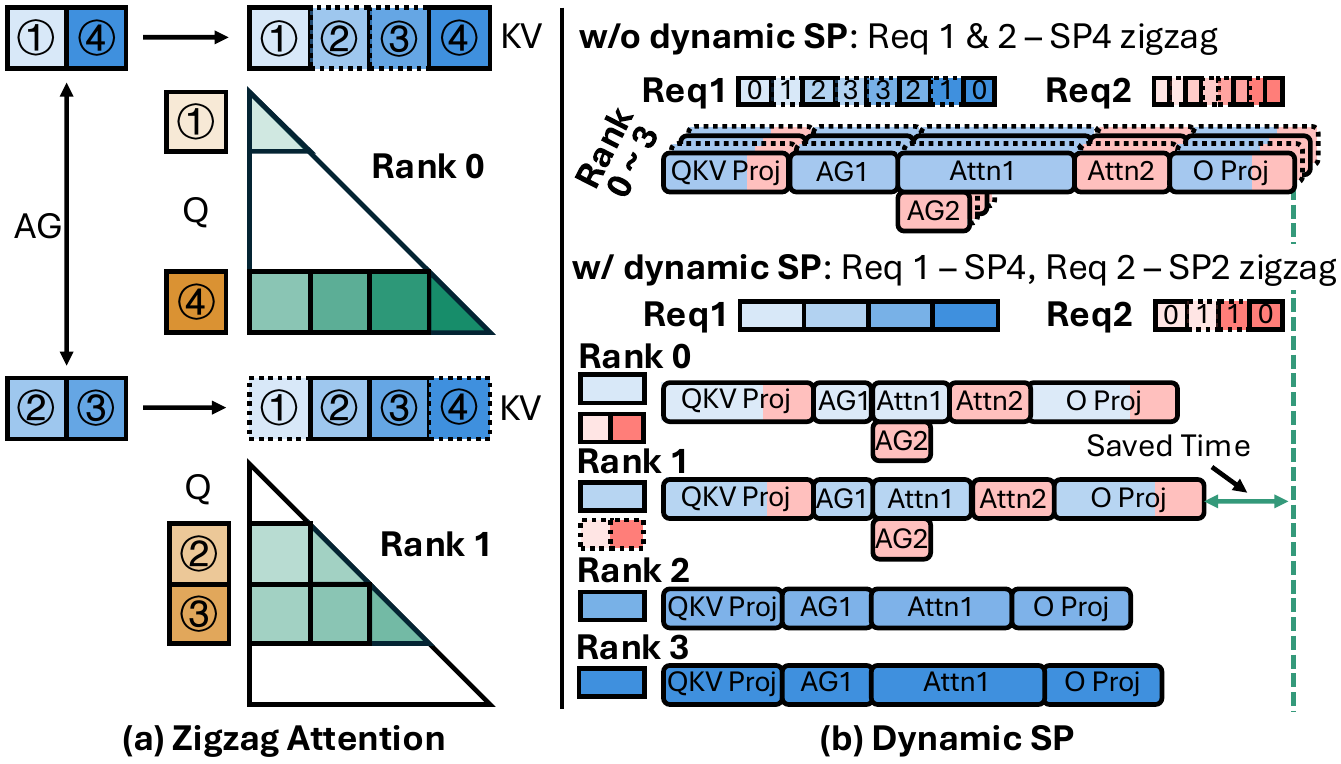}
    \caption{Illustration of zigzag attention and dynamic SP planning. Each request can adopt a different SP configuration: request 1 adopts SP4, and request 2 adopts SP2 with zigzag.}
    \label{fig:dynamic_sp}
    \vspace{-12pt}
\end{figure}

We used \charon~to analyze the tradeoffs between computation distribution and communication overhead under varying sequence length distributions. Unlike zigzag’s one-size-fits-all pattern, the simulator-generated dynamic SP plans assign different SP and zigzag configurations to each request within a batch. This fine-grained approach allows the system to balance compute and communication more precisely: for requests with long sequences, dynamic SP maintains zigzag-style balancing across ranks; for requests with shorter sequences, it reduces over-partitioning to avoid excessive all-gather overhead. As shown in Figure~\ref{fig:dynamic_sp}, dynamic SP adaptively adjusts the partitioning strategy per request, resulting in a more balanced utilization of GPU ranks across heterogeneous sequence lengths. By modeling both kernel execution and NCCL communication, \charon~predicts per-rank latency and generates optimal request-level SP strategies that minimize attention latency during the prefill stage.

We evaluated dynamic SP planning on LLaMA-3 70B model using 8x Nvidia Ada Lovelace inference GPUs and observed an average 15\% reduction in attention block latency compared to the zigzag baseline. The performance gains primarily stemmed from better handling of small-sequence requests, where communication overhead dominates. Specifically, disabling zigzag partitioning for short sequences avoided unnecessary all-gather costs, while mixing different SP plans across requests further balanced computation and communication. Together, these adjustments allowed the system to adapt SP strategies to workload characteristics, delivering consistent improvements over static planning. Importantly, we expect these gains to be larger on GPUs with higher communication latency (e.g., PCIe-based interconnects), where reducing communication overhead becomes even more critical.

\subsection{Optimal Inference Performance via Simulation}

LLM inference is a complex multi-objective optimization problem. It is governed by key performance metrics such as TTFT and TPOT, which together determine user-facing throughput, measured in Tokens Per Second per user (TPS/user). On the hardware side, system throughput, measured in Tokens Per Second per GPU (TPS/GPU), dictates the cost per token. These metrics are jointly influenced by numerous implementation choices, including scheduling algorithms and parallelism strategies (e.g., tensor and pipeline parallelism sizes, prefill chunk sizes, and batch sizes for prefill and decoding). A common deployment requirement is to maximize TPS/GPU (thereby minimizing cost) while adhering to user-facing performance constraints, such as specific Service Level Objectives (SLOs) for TTFT and TPOT. Deriving an optimal configuration analytically is often intractable due to the intricate interdependencies among these parameters. Consequently, the prevailing approach relies on extensive empirical benchmarking across various inference configurations. This method incurs significant time and monetary costs and requires deep expertise in how different parameter combinations interact—knowledge that most users lack. As a result, tuning an inference serving system to its optimal performance is a formidable challenge.

\begin{figure}[t]
    \centering
    \includegraphics[width=\linewidth]{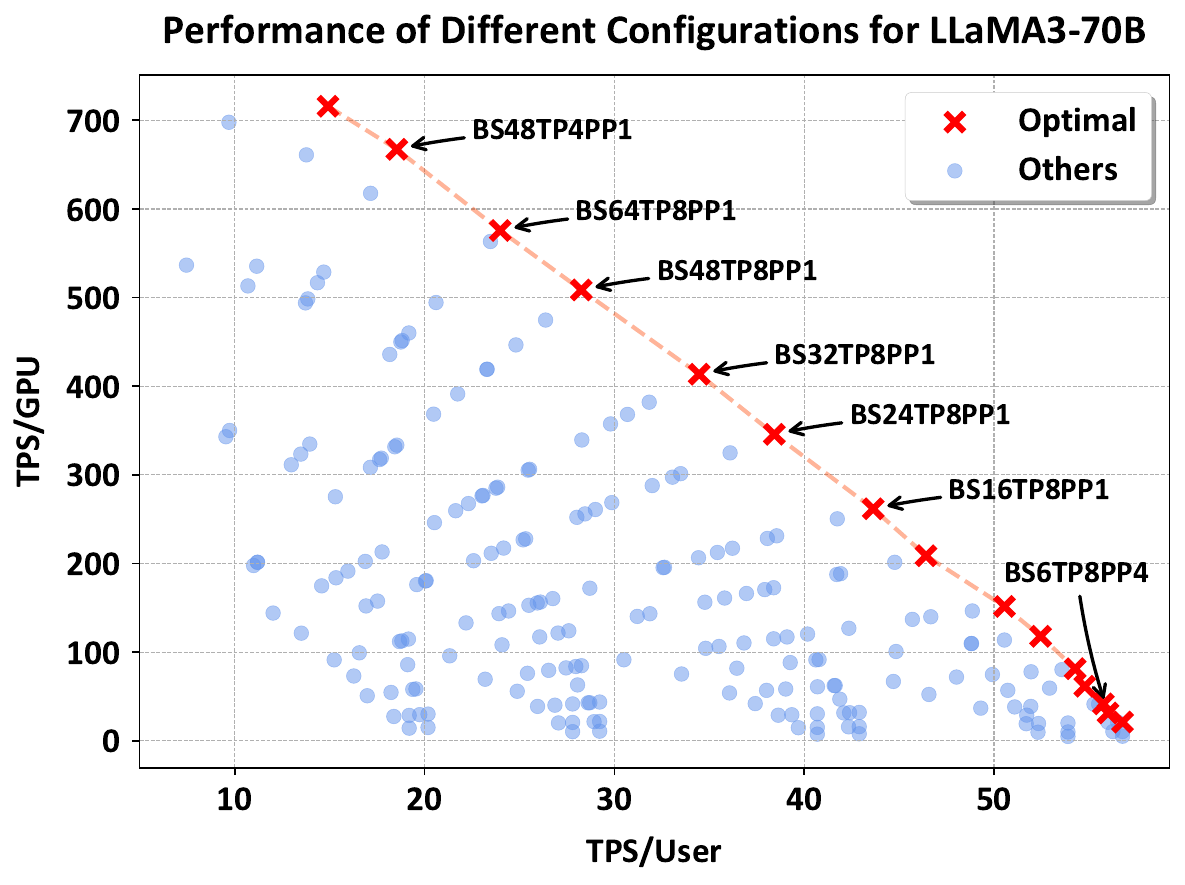}
    \vspace{-9pt}
    \caption{Simulated performance trade-off between system throughput (TPS/GPU) and user-facing throughput (TPS/User) for Llama 3 70B on NVIDIA Hopper Inference GPUs. Each point represents a unique inference configuration (TP, PP, batch sizes, etc.). The red dots form the optimal performance frontier identified by \charon, while blue dots represent sub-optimal configurations.}
    \label{fig:xllm}
    \vspace{-12pt}
\end{figure}

We address this challenge by enabling a comprehensive exploration of the cost-latency trade-off space at a negligible time cost, typically completing its analysis within two minutes. This allows for the rapid identification of optimal deployment parameters that satisfy given TPS/user constraints. As illustrated in Figure~\ref{fig:xllm} for a Llama 3 70B model, \charon~explores a vast space of configuration combinations. The optimal points (red) identified by \charon~offer a significant TPS/GPU improvement over sub-optimal configurations (blue) for the same user-facing performance. Furthermore, the plot reveals a clear trade-off along the optimal frontier: relaxing user-facing TPS constraints can yield up to a 7x increase in system throughput (TPS/GPU), thereby reducing costs.

To achieve its high speed, \charon~leverages pre-profiled kernel latencies and employs multi-process parallel simulation. For ease of use, it is deeply integrated with both our in-house inference framework and the standard Hugging Face Transformers library. \charon~automatically parses model architectures from Hugging Face configurations. From our in-house framework, it extracts the PyTorch FX graph representation and the corresponding kernel implementations used for inference. This design minimizes the effort required to use \charon~to nearly zero.

\charon~has been successfully validated in our production environment. In highly constrained deployments where manual tuning by generalist engineers often falls short, \charon~rapidly identifies optimal parameter combinations. For instance, when optimizing a fixed-output-length model service under a strict 100ms end-to-end latency SLO, \charon~discovered a configuration that vastly outperformed the manually tuned baseline. More importantly, as demonstrated by the Pareto frontier analysis in Figure~\ref{fig:xllm}, \charon~empowers engineers to systematically navigate the tradeoffs between user-facing latency and underlying hardware efficiency. By abstracting away the complexity of the configuration space, \charon~enables reliable maximum system throughput under specified SLOs without requiring extensive domain expertise.

%% file: sections/conslusion.tex
\section{Conclusion}

We present \charon, a unified, fine-grained simulator for large-scale LLM training and inference. \charon~achieves high fidelity through its compiler-style, graph-based architecture, enabling operator-level simulation to accurately model complex parallelism and communication-computation overlap. It simplifies workflows by natively ingesting PyTorch models and balances simulation cost and precision via its hybrid multi-engine backend. Our experiments demonstrate \charon~achieves an overall simulation error of less than 5.35\%, and under 3.74\% for a large-scale training. As a fast and accurate ``all-in-one'' platform, Charon significantly lowers the cost and expertise required to find optimal configurations for large-scale LLM training and serving deployment.

%% file: sections/ackownledgments.tex
\section{Acknowledgments}

We sincerely thank our teammates and colleagues for their continuous support and helpful suggestions during the development of Charon. Building this system required extensive teamwork, and we are deeply grateful to everyone who provided feedback along the way. We express our special thanks to Yan Xu, Chengji Yao, Xiang Li, Fan Yin, Dai Teng, Xiao Yu, and Xin Liu for their valuable ideas, project discussions, and code contributions. The successful completion of Charon would not have been possible without their tremendous support and collaboration.